# Fractional radial-cylindrical diffusivity model for levels of heterogeneity in petroleum reservoirs

George Parker-Lamptey[1,2] · Peter Amoako-Yirenkyi[1,2]  · Isaac Kwame Dontwi[1,2]

**Abstract** The generalized Taylor's Formula is used to derive a fractional radial-cylindrical diffusivity model via a fractional conservation of mass in radial geometry in a petroleum reservoir. The result is a space-fractional generalization of the diffusivity model with an arbitrary order that explains the degree of heterogeneity of the medium (a continuous spectrum of zero to one where zero heterogeneity equates to homogeneity). An implicit unconditionally stable numerical difference scheme of the linear form of the derived model is used to illustrate deviations of the model from the homogeneous medium. The variation of the order produces additional localized pressure drops during flow which is congruous with fluid inhibition effect of heterogeneities.

## Introduction

The theory of fractional calculus since its formulation in 1695 has in the last few decades been widely applied to a myriad of real life phenomena (Li et al. 2012). Fractional calculus is a branch of applied mathematics that generalizes classical differentiation and integration to calculus of arbitrary (non-integer) order. This concept practically extends the domain of the order of traditional models and thus allows the obtainment of more accurate results in modeling systems of complex structure. The advantage of using non-integer models lies in their ability to capture memory and hereditary processes of different substances (Podlubny 1998), a characteristic not exhibited by the classical integer order models.

Modeling with fractional calculus has proved to be both convincing and novel with extensive applications in several areas including physics (Barkai et al. 2000), systems biology (Yuste and Lindenberg 2001), biochemistry (Yuste et al. 2006), hydrology (Benson et al. 2000) and finance (Raberto et al. 2002). Some specific contributions of fractional calculus in modeling include the fractional modeling of heat transfer in heterogeneous media with the assumption that the heat is dispersed in the air around the beam (Sierociuk et al. 2013), anomalous diffusion studies in physics where particles spread at a rate inconsistent with the classical Brownian motion model (Pagnini et al. 2013) and applications in porous media transport among others.

The modeling of fluid flow in porous media within the petroleum reservoir features two basic equations: (1) the material balance or continuity equation and (2) the equation of motion (Darcy's Law). The combination of these equations usually with additional relationships results in partial differential equations of integer order. The continuity equation derived via the continuum approach relates the mass accumulation to the divergence (Bear and Cheng 2010). The divergence in this case is obtainable in the form of a first-order Taylor series representation with the assumption that the mass flux is linear. This assumption is however not appropriate for accounting for flux changes in large heterogeneous porous media (Wheatcraft and Meerschaert 2008). The problem occurs principally because of the variation of properties spatially (Dawe 2004;

✉ Peter Amoako-Yirenkyi
amoakoyirenkyi@knust.edu.gh

1 Department of Mathematics, Kwame Nkrumah University of Science and Technology, Kumasi, Ghana
2 Center for Scientific and Technical computing, National Institute for Mathematical Sciences, Kumasi, Ghana

Wheatcraft and Meerschaert 2008). Hydrocarbon reservoirs are typical examples of porous media which are heterogeneous in nature, yet the effects of heterogeneity are often not well represented and this becomes only obvious late during water production before the time predicted (Boyle et al. 2000) thus sparking several studies into the heterogeneity of hydrocarbon reservoirs.

A study by Wheatcraft and Meerschaert (2008) showed that by using the generalized Taylor series for approximating the mass flux, the limitations of the classical Taylor series are puzzled out thus resulting in the fractional conservation of mass model in rectilinear geometry. In this paper, we construct a radial-cylindrical diffusivity model for near wellbore single phase analysis which explains well the effect of heterogeneity in the hydrocarbon reservoir.

The paper is organized as follows. In "Fractional calculus" section is a presentation of some important concepts in fractional calculus. In "Mathematical modeling of single phase flow" section, the proposed model is presented with emphasis on the limitations of the existing model. We conclude with a presentation of the numerical stability, convergence and simulation of the proposed model and discussion of its novelty.

## Fractional calculus

There are several definitions formulated for fractional derivatives and integrals of which the most familiar are the Riemann-Liouville, Caputo and Grunwald-Letnikov definitions which customarily emerge from fractional order integrals, derivatives and differences respectively (Monje et al. 2010). The Riemann-Liouville and Caputo definitions and their differences landmarks their fields of application. Engineers for example prefer the Caputo derivatives primarily due to its uniformity of results with the integer order differential equations, whereas Mathematicians and Physicists usually apply the Riemann–Liouville (Podlubny 1998). In this paper, we use the Caputo definition for its advantage of ensuring that initial conditions of fractional differential equations are cognate to those of integer order differential equations (Caputo 2012) and uniformity with the use of Generalized Taylor's series.

**Definition 1** The Gamma function $\Gamma(x)$ is defined as

$$\Gamma(x) = \int_0^\infty e^{-t} t^{x-1} dt \quad (1)$$

and is convergent for $x > 0$.

**Definition 2** The $\alpha$th Caputo fractional derivative of the function $f(t)$ is defined as

$$_0D_t^\alpha f(t) = \frac{1}{\Gamma(n-\alpha)} \int_0^t \frac{f^{(n)}(\tau)}{(t-\tau)^{\alpha-n+1}} d\tau \quad (2)$$

where $n-1 < \alpha < n$ such that $n \in \mathbb{Z}_+$.

**Definition 3** The *Fractional Taylor's Series* is the generalization of the classical Taylor Series. For $\alpha$ as the order of differentiation, suppose $D_a^{k\alpha} f(x) \in C(a,b]$ for $k = 0, 1, \ldots, m+1$ where $0 < \alpha \leq 1$, the Fractional Taylor Series is defined as

$$f(x) = \sum_{n=0}^m \frac{(x-a)^{n\alpha}}{\Gamma(n\alpha+1)} \left(D_a^{n\alpha} f\right)(a) + \frac{\left(D_a^{(m+1)\alpha}\right)(\xi)}{\Gamma((m+1)\alpha+1)} \cdot (x-a)^{(n+1)\alpha} \quad (3)$$

where $a \leq \xi \leq x \ \forall x \in (a,b]$ (Odibat and Shawagfeh 2007).

## Mathematical modeling of single phase flow

### Introduction

The governing equations for a single phase flow evolves from a combination of fundamental concepts i.e., the conversation of mass, conservation of momentum and an equation of state for the porous medium when considered as a control volume. The shape of the control volume with which the system is modeled depends on the coordinate system which should conform as closely as possible to the flow geometry. The form of geometries in petroleum reservoir modeling is either rectilinear or curvilinear (Turgay et al. 2001), the latter, features the cylindrical coordinate exclusively used in single-well simulation problems (Ahmed 2006). Based on the cylindrical geometry displayed in Fig. 1, the control volume with sides $\Delta r$, $\Delta \theta$ and $\Delta z$ and with initial point $(r, \theta, z)$, the mass flux through the $r$-(1265) face is

$$F(r) = r\Delta\theta\Delta z(\varsigma_c \rho u_r) \quad (4)$$

where $\varsigma_c$ = volume conservation factor, $\rho$ = fluid density and $u_r$ = volumetric velocity in the $r$-direction. The mass outflow at the $r + \Delta r$ end of the control volume is obtained by summing the mass flux through the $r$-face and the change that take place in the $r$-direction multiplied by the length $\Delta r$ over which the change acts. This results in

$$F(r+\Delta r) = \varsigma_c \Delta\theta\Delta z(r\rho u_r) + \varsigma_c \Delta\theta\Delta z \frac{\partial(r\rho u_r)}{\partial r} \Delta r$$
$$= \varsigma_c \Delta\theta\Delta z \left[r\rho u_r + \frac{\partial(r\rho u_r)}{\partial r} \Delta r\right] \quad (5)$$

The relation in (5) is the Taylor Series expansion about the point $r$ of the first order of which the exact expansion is given by

$$F(r + \Delta r) = \varsigma_c \Delta\theta\Delta z \left( r\rho u_r + \frac{\partial(r\rho u_r)}{\partial r}\Delta r \right.$$
$$\left. + \frac{\partial^2(r\rho u_r)}{\partial r^2}\frac{(\Delta r)^2}{2!} + \cdots \right) \quad (6)$$

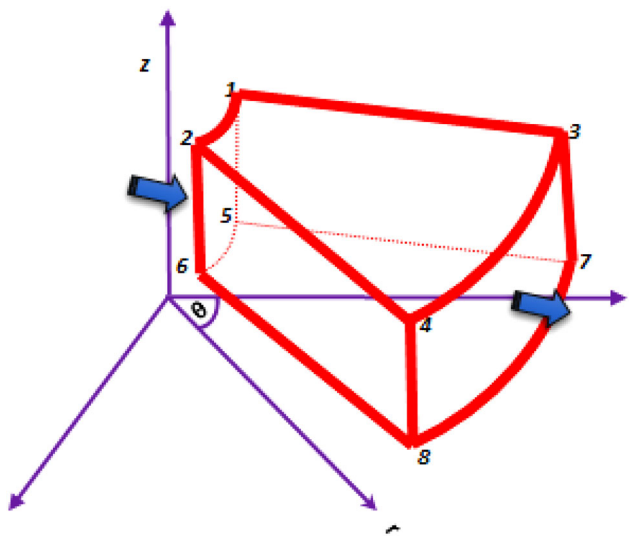

**Fig. 1** Control volume in cylindrical coordinates

Equation (6) is the classical Taylor series expansion about the point $r$ and the assumption of using (5) as a form of approximation for (6) is that changes in the mass flux are linear or piecewise linear if considered as nonlinear.

The expression representing the difference between the mass entering the volume at $r$ and the mass leaving at $r + \Delta r$ is given by

$$F(r) - F(r + \Delta r) = \varsigma_c \Delta\theta\Delta z(r\rho u_r)$$
$$- \varsigma_c \Delta\theta\Delta z\left(r\rho u_r + \frac{\partial(r\rho u_r)}{\partial r}\Delta r\right) \quad (7)$$
$$= -\varsigma_c \frac{V_b}{r}\frac{\partial(r\rho u_r)}{\partial r}$$

where the bulk volume of the control element, $V_b = r\Delta r\Delta\theta\Delta z$.

Similar to the derivation for the mass exchange in the radial component within the element, the mass exchange in the $\theta$-direction is given by

$$F(\theta) - F(\theta + \Delta\theta) = \varsigma_c \Delta r\Delta z(\rho u_\theta)$$
$$- \varsigma_c \Delta r\Delta z\left(\rho u_\theta + \frac{\partial(\rho u_\theta)}{\partial\theta}\Delta\theta\right) \quad (8)$$
$$= -\varsigma_c \frac{V_b}{r}\frac{\partial(\rho u_\theta)}{\partial\theta}$$

Also the $z$-directional component is equally given by

$$F(z) - F(z + \Delta z) = \varsigma_c r\Delta r\Delta\theta(\rho u_z)$$
$$- \varsigma_c r\Delta r\Delta\theta\left(\rho u_z + \frac{\partial(\rho u_z)}{\partial z}\Delta z\right) \quad (9)$$
$$= -\varsigma_c V_b \frac{\partial(\rho u_z)}{\partial z}$$

By equating the sum of the divergence in $r, \theta$ and $z$ coordinates to the mass accumulation term, i.e., $V_b\partial(\phi\rho)/\partial t$, and dividing through the equation by $\varsigma_c V_b$, the result is the mass-balance equation of the form

$$-\left[\frac{1}{r}\frac{\partial(r\rho u_r)}{\partial r} + \frac{1}{r}\frac{\partial(\rho u_\theta)}{\partial\theta} + \frac{\partial(\rho u_z)}{\partial z}\right] = \frac{1}{\varsigma_c}\frac{\partial(\phi\rho)}{\partial t} \quad (10)$$

where $\phi$ = porosity. Equation (10) is the three dimensional radial-cylindrical mass-balance equation without external sink/source. This is because the well is assumed to be at the center of the drainage area and the external conditions are incorporated in the form of boundary conditions (Turgay et al. 2001).

### Fractional radial-cylindrical diffusivity model

In order to remove the restriction that the flux has to be linear and the control volume has to be infinitesimal (Wheatcraft and Meerschaert 2008), we replace the classical Taylor series approximation for the divergence in each coordinate $r, \theta$ and $z$ with the fractional first-order Taylor series approximation of (3). The component of the flux through the $r + \Delta r$ section of the radial-cylindrical control volume is given by

$$F(r + \Delta r) = \varsigma_c \Delta\theta\Delta z(r\rho u_r) + \varsigma_c \Delta\theta\Delta z \frac{\partial^{\alpha_r}(r\rho u_r)}{\partial r^{\alpha_r}}\frac{\Delta r^{\alpha_r}}{\Gamma(\alpha_r + 1)} \quad (11)$$

This implies that the net flux in the $r$-direction is given by

$$F(r) - F(r + \Delta r) = -\varsigma_c \Delta\theta\Delta z \frac{\partial^{\alpha_r}(r\rho u_r)}{\partial r^{\alpha_r}}\frac{\Delta r^{\alpha_r}}{\Gamma(\alpha_r + 1)} \quad (12)$$

The net mass flux in the $\theta$ and $z$ directions are given by (13) and (14) respectively. For an anisotropic medium, the fractional orders are not the same in all directions. Therefore, we have associate $\alpha_\theta$-order and $\alpha_z$-order to the $\theta$-direction and $z$-direction.

$$F(\theta) - F(\theta + \Delta\theta) = -\varsigma_c \Delta r\Delta z \frac{\partial^{\alpha_\theta}(\rho u_\theta)}{\partial\theta^{\alpha_\theta}}\frac{\Delta\theta^{\alpha_\theta}}{\Gamma(\alpha_\theta + 1)} \quad (13)$$

$$F(z) - F(z + \Delta z) = -\varsigma_c r\Delta r\Delta\theta \frac{\partial^{\alpha_z}(\rho u_z)}{\partial z^{\alpha_z}}\frac{\Delta z^{\alpha_z}}{\Gamma(\alpha_z + 1)} \quad (14)$$

The radial fractional mass-balance model formulated by equating the net mass flux through the control volume to the mass accumulation term is obtained as

$$-V_b\varsigma_c \left[ \frac{1}{r} \frac{\Delta r^{\alpha_r-1}}{\Gamma(\alpha_r+1)} \frac{\partial^{\alpha_r}(r\rho u_r)}{\partial r^{\alpha_r}} + \frac{1}{r} \frac{\Delta \theta^{\alpha_\theta-1}}{\Gamma(\alpha_\theta+1)} \frac{\partial^{\alpha_\theta}(\rho u_\theta)}{\partial \theta^{\alpha_\theta}} \right.$$
$$\left. + \frac{\Delta z^{\alpha_z-1}}{\Gamma(\alpha_z+1)} \frac{\partial^{\alpha_z}(\rho u_z)}{\partial z^{\alpha_z}} \right] = \frac{\partial(V_b \phi \rho)}{\partial t} \quad (15)$$

Equation (15) represents the three dimensional radial-cylindrical fractional conservation of mass or fractional continuity equation for a heterogeneous but anisotropic medium without external source/sink terms. It is important to note here that when the fractional orders are individually equal to one, i.e., $\alpha_r = \alpha_\theta = \alpha_z = 1$, we obtain the classical three dimensional conservation of mass equation for radial geometry.

The second phase in the construction of the fluid flow model for the petroleum reservoir involves the introduction of conservation of momentum (Darcy's law) in radial-cylindrical coordinates. The relations for the superficial-velocity components in the $r, \theta,$ and $z$ directions assuming that the potential gradient is equivalent to the pressure gradient are given by

$$u_r = -\beta_c \frac{k_r}{\mu}\frac{\partial p}{\partial r}, \qquad u_\theta = -\beta_c \frac{k_\theta}{\mu}\frac{1}{r}\frac{\partial p}{\partial \theta}, \qquad u_r = -\beta_c \frac{k_z}{\mu}\frac{\partial p}{\partial z} \quad (16)$$

where $\beta_c$—transmissibility conversion factor, $k_i$— permeability in the direction of $i$, $p$—fluid pressure and $\mu$—fluid viscosity.

By substituting the expressions for the fluid components in (16) into (15) and setting the fluid density, $\rho = \rho_{sc}/B$ where $\rho_{sc}$ is the fluid density at standard conditions and $B$ is the formation volume factor, we obtain

$$V_b\varsigma_c \left[ \frac{1}{r} \frac{\Delta r^{\alpha_r-1}}{\Gamma(\alpha_r+1)} \frac{\partial^{\alpha_r}}{\partial r^{\alpha_r}}\left(r\beta_c \frac{k_r}{\mu B}\frac{\partial p}{\partial r}\right) \right.$$
$$+ \frac{1}{r^2} \frac{\Delta \theta^{\alpha_\theta-1}}{\Gamma(\alpha_\theta+1)} \frac{\partial^{\alpha_\theta}}{\partial \theta^{\alpha_\theta}}\left(\beta_c \frac{k_\theta}{\mu B}\frac{\partial p}{\partial \theta}\right)$$
$$\left. + \frac{\Delta z^{\alpha_z-1}}{\Gamma(\alpha_z+1)} \frac{\partial^{\alpha_z}}{\partial z^{\alpha_z}}\left(\beta_c \frac{k_z}{\mu B}\frac{\partial p}{\partial z}\right) \right] = \frac{\partial}{\partial t}\left(\frac{V_b \phi}{B}\right) \quad (17)$$

We now turn our attention to the mass accumulation term on the right-hand side of (17). We assume here that there is only vertical compressibility in the control volume, i.e. the control volume changes over time only in the vertical dimension, $\Delta z$. We can then express the mass accumulation term as

$$\frac{\partial}{\partial t}\left(\frac{V_b \phi}{B}\right) = r\Delta r \Delta \theta \left[ \frac{\phi}{B}\frac{\partial}{\partial t}(\Delta z) + \Delta z \phi \frac{\partial}{\partial t}\left(\frac{1}{B}\right) + \frac{\Delta z}{B}\frac{\partial}{\partial t}(\phi) \right] \quad (18)$$

The first term in (18) which has time derivative of the change in the vertical dimension is expressed as

$$\frac{\partial}{\partial t}(\Delta z) = \frac{\partial}{\partial p}(\Delta z)\frac{\partial p}{\partial t} = \beta_s \Delta z \frac{\partial p}{\partial t} \quad (19)$$

where $\beta_s$ is the coefficient of compressibility for the porous medium same as obtained by Wheatcraft and Meerschaert (2008) for linear geometry. Also we consider the second term in (18) by replacing $(1/B) = [1 + c(p - p^o)]/B^o$ where $B^o$ and $p^o$ represent the formation volume factor and pressure at standard conditions respectively and $c$ is the fluid compressibility. The time derivative of $B^{-1}$ is thus given by

$$\frac{\partial}{\partial t}\left(\frac{1}{B}\right) = \frac{\partial}{\partial t}\left(\frac{[1+c(p-p^o)]}{B^o}\right) = \frac{c}{B^o}\frac{\partial p}{\partial t} \quad (20)$$

The third term in (18) involving the time derivative of the porosity is similarly simplified using the relation $\phi = \phi^o[1 + c_R(p - p^o)]$ where $\phi^o$ is the porosity at standard conditions and $c_R$ is the rock compressibility. Therefore

$$\frac{\partial \phi}{\partial t} = \frac{\partial \phi}{\partial p}\frac{\partial p}{\partial t} = \frac{\partial}{\partial p}\{\phi^o[1+c_R(p-p^o)]\}\frac{\partial p}{\partial t} = c_R \phi^o \frac{\partial p}{\partial t} \quad (21)$$

The mass accumulation term can be simplified by substituting (19), (20) and (21) into (18) resulting in

$$\frac{\partial}{\partial t}\left(\frac{V_b \phi}{B}\right) = \frac{V_b \phi}{B^o}\{\beta_s[1+c(p-p^o)] + c + c_R \frac{\phi^o}{\phi}[1+c(p-p^o)]\}\frac{\partial p}{\partial t} \quad (22)$$

Since the compressibility, $c$ is small ($10^{-5} - 10^{-6}\ psi^{-1}$) (Turgay et al. 2001) and $c_R$ is equally small, the mass accumulation can be expressed as

$$\frac{\partial}{\partial t}\left(\frac{V_b \phi}{B}\right) = \frac{V_b \phi}{B^o}(\beta_s + c_t)\frac{\partial p}{\partial t} \quad (23)$$

where $c_t$ is the total compressibility of the fluid and the formation, i.e. $c_t = c + c_R \phi^o/\phi$.

Substituting (23) into (17) we have

$$\frac{1}{r}\frac{\Delta r^{\alpha_r-1}}{\Gamma(\alpha_r+1)}\frac{\partial^{\alpha_r}}{\partial r^{\alpha_r}}\left(r\beta_c \frac{k_r}{\mu B}\frac{\partial p}{\partial r}\right) + \frac{1}{r^2}\frac{\Delta \theta^{\alpha_\theta-1}}{\Gamma(\alpha_\theta+1)}\frac{\partial^{\alpha_\theta}}{\partial \theta^{\alpha_\theta}}\left(\beta_c \frac{k_\theta}{\mu B}\frac{\partial p}{\partial \theta}\right)$$
$$+ \frac{\Delta z^{\alpha_z-1}}{\Gamma(\alpha_z+1)}\frac{\partial^{\alpha_z}}{\partial z^{\alpha_z}}\left(\beta_c \frac{k_z}{\mu B}\frac{\partial p}{\partial z}\right) = \frac{\phi}{\varsigma_c B^o}(\beta_s + c_t)\frac{\partial p}{\partial t} \quad (24)$$

Equation (24) is the fractional radial-cylindrical diffusivity model for a heterogeneous medium with allowance for vertical compressibility of the control volume with each coordinate having an independent order. The classical radial-cylindrical diffusivity model with vertical compressibility is obtained if the fractional orders are equal to 1, i.e. $\alpha_r = \alpha_\theta = \alpha_z = 1$ as

$$\frac{1}{r}\frac{\partial}{\partial r}\left(r\beta_c \frac{k_r}{\mu B}\frac{\partial p}{\partial r}\right) + \frac{1}{r^2}\frac{\partial}{\partial \theta}\left(\beta_c \frac{k_\theta}{\mu B}\frac{\partial p}{\partial \theta}\right) + \frac{\partial}{\partial z}\left(\beta_c \frac{k_z}{\mu B}\frac{\partial p}{\partial z}\right)$$
$$= \frac{\phi}{\varsigma_c B^o}(\beta_s + c_t)\frac{\partial p}{\partial t} \quad (25)$$

The fractional radial-cylindrical diffusivity model in one dimension ($r$ direction) for a slightly compressible liquid in a heterogeneous medium for $\alpha_r = \alpha$ is given by

$$\frac{1}{r}\frac{\Delta r^{\alpha-1}}{\Gamma(\alpha+1)}\frac{\partial^\alpha}{\partial r^\alpha}\left(r\beta_c\frac{k_r}{\mu B}\frac{\partial p}{\partial r}\right) = \frac{\phi}{\varsigma_c B^o}(\beta_s + c_t)\frac{\partial p}{\partial t} \quad (26)$$

For a constant permeability reservoir, $k_r$ = constant, $\mu$ is constant for slightly compressible liquids. We replace $B^{-1} = [1 + c(p - p^0)]/B^0$ resulting in

$$\frac{1}{r}\frac{\Delta r^{\alpha-1}}{\Gamma(\alpha+1)}\left\{\frac{\partial^\alpha}{\partial r^\alpha}\left(r\frac{\partial p}{\partial r}\right)[1+c(p-p^0)] + c\left(\frac{\partial^\alpha p}{\partial r^\alpha}\right)\left(\frac{\partial p}{\partial r}\right)\right\}$$
$$= \frac{1}{\chi}\frac{\partial p}{\partial t} \quad (27)$$

where $\chi = (k\varsigma_c\beta_c)/(\phi\mu(\beta_s + c_t))$. For small compressibility $c$, $[1 + c(p - p^0)] \approx 1$ and the nonlinear term $c(\partial^\alpha p/\partial r^\alpha)(\partial p/\partial r)$ is negligible and thus the linear fractional diffusivity model in one-dimensional radial geometry is

$$\frac{1}{r}\frac{\Delta r^{\alpha-1}}{\Gamma(\alpha+1)}\frac{\partial^\alpha}{\partial r^\alpha}\left(r\frac{\partial p}{\partial r}\right) = \frac{1}{\chi}\frac{\partial p}{\partial t} \quad (28)$$

Evaluating the Caputo derivatives and simplifying (28), we obtain

$$\frac{1}{r^\alpha}\frac{\Delta r^{\alpha-1}}{\Gamma(\alpha+1)\Gamma(2-\alpha)}\frac{\partial p}{\partial r} + \frac{\Delta r^{\alpha-1}}{\Gamma(\alpha+1)}\frac{\partial^\alpha}{\partial r^\alpha}\left(\frac{\partial p}{\partial r}\right) = \frac{1}{\chi}\frac{\partial p}{\partial t} \quad (29)$$

## Fractional numerical approximation

To establish the numerical solution of the linear fractional radial-cylindrical model, let $t_k = k\tau (k = 0, 1, 2, \ldots, n)$ be the discretized time domain such that $t_k \in [0, T]$, $\tau = \frac{T}{n}$. The radial coordinates are logarithmically spaced away from the wellbore where $r \in [r_w, r_e]$ where $r_e$—radius of well and $r_w$—radius of external boundary. The discretized space domain is $r_{i+1} = \varphi r_i$ ($i = 1, 2, \ldots, n_r - 1$) where $\varphi = \left(\frac{r_e}{r_w}\right)^{1/(n_r-1)}$ and $h_i = \Delta r_i$. Let $p_i^k = p(r_i, t_k)$ be the numerical estimate of the exact solution of the pressure depending on time and space, $p(r, t)$ at the mesh point $(r_i, t_k)$. The finite difference approximation for the time derivative is the first forward difference in (29):

$$\frac{\partial p(r_i, t_k)}{\partial t} = \frac{p(r_i, t_{k+1}) - p(r_i, t_k)}{\tau_j} + O(\tau) \quad (30)$$

The first-order derivative in the $r$ direction is given central difference scheme

$$\frac{\partial p(r_i, t_{k+1})}{\partial r} = \frac{p(r_{i+1}, t_{k+1}) - p(r_{i-1}, t_{k+1})}{2h_i} + O(h)^2 \quad (31)$$

The Caputo derivative is discretized using for any arbitrary order $\gamma$ (Kilbas et al. 2006):

$$\frac{\partial^\gamma p(r,t)}{\partial r^\gamma} = ({}_0D_r^\gamma p)(r,t) - \frac{p(0,t)}{\Gamma(1-\gamma)}(r_i)^{-\gamma} + \frac{p'(0,t)}{\Gamma(2-\gamma)}(r_i)^{1-\gamma} \quad 1 < \gamma \leq 2 \quad (32)$$

where $({}_0D_r^\gamma p)(r,t)$ is the Riemann-Liouville fractional partial derivative of order $\gamma \in (1, 2]$. The Riemann-Liouville derivative is discretized using the Grunwald formula (Podlubny 1998):

$$({}_0D_r^\gamma p)(r,t) = \frac{1}{h_i^\gamma}\sum_{j=0}^{i+1} g_j p(r_{i-j+1}, t_{k+1}) + O(h) \quad (33)$$

where $g_0 = 1$, $g_1 = -\gamma$, $g_n = (-1)^n \binom{\gamma}{n}$ represents the normalized Grunwald weights. By discretizing $p'(0, t)$ in (32) using the first forward difference scheme (Afshari et al. 2015), the Caputo space derivative of order $\gamma$ results in

$$\frac{\partial^\gamma p(r,t)}{\partial r^\gamma} = \frac{1}{h_i^\gamma}\sum_{j=0}^{i+1} g_j p(r_{i-j+1}, t_{k+1}) - \frac{p(r_1, t_{k+1})}{h_i\Gamma(2-\gamma)}(r_i)^{1-\gamma} + \left[\frac{(r_i)^{1-\gamma}}{h_i\Gamma(2-\gamma)} - \frac{(r_i)^{-\gamma}}{\Gamma(1-\gamma)}\right]p(r_0, t_{k+1}) + O(h) \quad (34)$$

where $1 < \gamma \leq 2$.

Substituting the derivative approximations in (30), (31) and (34) into (29), the result is the implicit difference scheme with $\gamma = \alpha + 1$:

$$\Psi_{1i}\left(p_{i+1}^{k+1} - p_{i-1}^{k+1}\right) + \Psi_{2i}\sum_{j=0}^{i+1} g_j p_{i-j+1}^{k+1} - \Psi_{3i} p_1^{k+1} + \Psi_{4i} p_0^{k+1} - p_i^{k+1} = -p_i^k \quad (35)$$

for $i = 1, 2, \ldots, m$ and $k = 0, 1, 2, \ldots, n$

$$\Psi_{1i} = \frac{\tau_k \chi_{ij}^k}{\Gamma(\alpha+1)}\frac{h_i^{\alpha-2}}{2\Gamma(2-\alpha)r_i^\alpha}, \qquad \Psi_{2i} = \frac{\tau_k \chi_{ij}^k}{\Gamma(\alpha+1)}\frac{1}{h_i^{1+\gamma-\alpha}},$$

$$\Psi_{3i} = \frac{\tau_k \chi_{ij}^k}{\Gamma(\alpha+1)}\frac{(r_i)^{1-\gamma}}{\sigma h_1 h_i^{1-\alpha}\Gamma(2-\gamma)},$$

$$\Psi_{4i} = \frac{\tau_k \chi_{ij}^k h_i^{\alpha-1}}{\Gamma(\alpha+1)}\left[\frac{(r_i)^{1-\gamma}}{h_1\Gamma(2-\gamma)} - \frac{(r_i)^{-\gamma}}{\Gamma(1-\gamma)}\right]$$

The system in (35) can be written in matrix form as

$$[\mathbf{A}_0\mathbf{A}_1 + \mathbf{A}_2 + \mathbf{A}_3 + \mathbf{A}_4]\mathbf{p}^{k+1} = -\mathbf{p}^k - [\mathbf{F}_1 + \mathbf{F}_2] \quad (36)$$

where $\mathbf{p}^k = [p_1^k, p_2^k, \ldots, p_{m-1}^k]^T$, $\mathbf{F}_1 = [0, 0, \ldots, 0, (\Psi_{1,m-1} + g_0\Psi_{2,m-1})]^T p_m^k$, $\mathbf{F}_2 = (\mathbf{F}_{21} + \mathbf{F}_{22} + \mathbf{F}_{23})p_0^k$ such that $\mathbf{F}_{21} = [-\Psi_{1,1}, 0, \ldots, 0]^T$, $\mathbf{F}_{22} = [g_2\Psi_{2,1}, g_3\Psi_{2,2}, \ldots, g_m$

$\Psi_{2,m}]^T$, $F_{23} = [\Psi_{4,1}, \Psi_{4,2}, \ldots, \Psi_{4,m-1}]^T$, $A_i$s are $(m-1) \times (m-1)$ matrices such that

$$A_0 = \begin{bmatrix} \Psi_{1,2} & 0 & \cdots & 0 \\ 0 & \Psi_{1,3} & \cdots & \vdots \\ \vdots & \ddots & \ddots & 0 \\ 0 & \cdots & \cdots & \Psi_{1,m-1} \end{bmatrix},$$

$$A_1 = \begin{bmatrix} g_1 & g_0 & 0 & \cdots & 0 \\ g_2 & g_1 & g_0 & \cdots & \vdots \\ \vdots & \vdots & \ddots & \ddots & g_0 \\ g_{m-1} & \cdots & \cdots & g_2 & g_1 \end{bmatrix}$$

$$A_2 = -\begin{bmatrix} \Psi_{3,1} & 0 & \cdots & 0 \\ \Psi_{3,2} & 0 & \cdots & \vdots \\ \vdots & \ddots & \vdots & 0 \\ \Psi_{3,m-1} & \cdots & 0 & 0 \end{bmatrix},$$

$$A_3 = \begin{bmatrix} 0 & \Psi_{1,1} & \cdots & 0 \\ -\Psi_{1,2} & 0 & \ddots & \vdots \\ 0 & \ddots & \ddots & \Psi_{1,m-2} \\ 0 & \cdots & -\Psi_{1,m-1} & 0 \end{bmatrix}$$

Also, $A_4 = -I_{(m-1),(m-1)}$ and $A_1$ is the lower triangular strip matrix.

### Stability

**Theorem 1** *The implicit numerical system defined by the linear difference equation in* (35) *has a unique and unconditionally stable solution.*

We prove the uniqueness of the solution via the application of the Gerschgorin theorem, every eigenvalue of the matrix $A = \{a_{ij}\}$ associated with (35) defined by

$$A = \begin{cases} g_0\Psi_{2,i} + \Psi_{1,i} & j = i+1 \\ g_1\Psi_{2,i} - \Psi_{3,i} & j = i = 1 \\ g_1\Psi_{2,i} - 1 & j = i, j \neq 1 \\ g_i\Psi_{2,i} - \Psi_{1i} - \Psi_{3,i} & j = 1, j \neq i, i = 2 \\ g_i\Psi_{2,i} - \Psi_{3,i} & j = 1, j \neq i, i \neq 2 \\ g_2\Psi_{2,i} - \Psi_{1,i} & j = i-1, j \neq 1 \\ g_{i-j+1}\Psi_{2i} & 1 < j \leq i-2 \end{cases} \quad (37)$$

has a magnitude greater than unity.

Let $\lambda$ be an eigenvalue of the matrix $A$. Based on the Grunwald weights, then for $1 < \gamma \leq 2$, the results indicate that $g_j \geq 0$ for $j \geq 2$. Also for any $v$,

$$(1+z)^v = \sum_{m=0}^{\infty} \binom{v}{m} z^m, \quad |z| \leq 1 \quad (38)$$

If $z = -1$ is substituted into (38), it can be seen that $\sum_{m=0}^{\infty} g_n = 0$ with $g_1 = -1$ being the only negative term in the sequence of the Grunwald weights. This means that $-g_1 = \gamma$ is the largest estimate and $-g_1 \geq \sum_{m=0, m\neq 1}^{k} g_m$ for any $k = 0, 1, 2, \ldots$. Based on the Gerschgorin theorem, all eigenvalues of $A = \{a_{ij}\}$ are contained in $n$ disks, $D_i = D_i(a_{ii}, s_i)$. The $a_{ii}$'s are given by

$$a_{ii} = \Psi_{3,i} - g_1\Psi_{2,i} + 1 = \Psi_{3,2} + \gamma\Psi_{2,1} + 1 \quad \text{for} \quad i = 1$$
$$a_{ii} = -g_1\Psi_{2,i} + 1 = \gamma\Psi_{2,1} + 1 \quad \text{for} \quad i \neq 1$$

and the radii $s_i$'s are computed by

$$s_i = \sum_{j=1, j\neq i}^{m-1} |a_{ij}| = \sum_{j=1, j\neq i}^{i+1} |a_{ij}|$$
$$= |g_0\Psi_{2,i} - \Psi_{1,i}| + \sum_{j=1}^{i-2} |-g_{i-j+1}\Psi_{2,i} + \Psi_{3,i}|$$
$$+ |-g_2\Psi_{2,i} + \Psi_{1,i}|$$
$$\leq \Psi_{2,i} + 2(\Psi_{2,i} + \Psi_{1,i}) + \Psi_{3,i}$$

Since $\Psi_{ij} \geq 0$ and from both cases of $a_{ii}$, it implies that every eigenvalue $\lambda$ of the matrix has a real part great than 1. Therefore, the spectral radius of the inverse matrix $\rho(A) \leq 1$. This means that the numerical scheme has unique solution.

We use the fractional Von Neumann stability procedure to show the unconditional stability of (35) by letting $p_i^k = \xi_n e^{\mathbf{I}qih}$ where $\mathbf{I} = \sqrt{-1}$, $q$ real. By substituting this into (35), we obtain

$$\Psi_{1,i}\left(e^{\mathbf{I}q(i+1)h} - e^{\mathbf{I}q(i-1)h}\right)\xi_{n+1}$$
$$- \Psi_{2,i}\sum_{j=0}^{i+1} g_j\xi_{n+1}e^{\mathbf{I}q(i-j+1)h} - \Psi_{3,i}\xi_{n+1}e^{\mathbf{I}qh} \quad (39)$$
$$- \Psi_{4,i}\xi_{n+1}e^{\mathbf{I}q0h} - \xi_{n+1}e^{\mathbf{I}qih} = -\xi_n e^{\mathbf{I}qih}$$

By dividing through (39) by $e^{\mathbf{I}qih}$ and further simplification, we obtain

$$\xi_{n+1} = \frac{\xi_n}{\left| -\Psi_{1,i}(e^{\mathbf{I}qh} - e^{-\mathbf{I}qh}) - \Psi_{2,i}\sum_{j=0}^{i+1} g_j e^{\mathbf{I}q(1-j)h} + \Psi_{3,i} e^{\mathbf{I}q(1-i)h} - \Psi_{4,i} e^{\mathbf{I}iqh} + 1 \right|} \quad (40)$$

By substituting Euler's formula $e^{\mathbf{I}\theta} = \cos\theta \pm \mathbf{I}\sin\theta$, the amplification factor $\mu_{max} = |\xi_{n+1}|/|\xi_n|$ is given by

$$\mu_{max} = \frac{1}{\left| -\Psi_{2,i}\sum_{j=0}^{i+1} g_j \cos q(j-1)h + \Psi_{3,i}\cos q(i-1)h + \Psi_{4,i}\cos qih + 1 \right|} \quad (41)$$

Using the properties of the grundwald weights, that is, $\sum_{k=0}^{\infty} g_j = 0$ and $\sum_{k=0}^{m} g_j < 0$, it means that the denominator in (41) is simplified to $\Psi_{3,i}^2 \cos^2 q(i-1)h + 2\Psi_{3,i}(\Psi_{4,i}\cos q(i-1)h + 1) + (\Psi_{4,i}\cos q(i-1)h + 1)^2$. Therefore $\mu_{max} \leq 1$ and hence the scheme is unconditionally stable.

## Numerical example

In this section, a numerical example of the proposed model in one dimension is presented. We consider a reservoir with properties as given in Table 1.

In addition we use a logarithmically spaced radial coordinate with wellbore radius $r_w = 0.25$ ft and a drainage radius $r_e = 2500$ ft. A constant initial pressure $p(r, t = 0) = 4000$ psi and a Neumann problem using the specified flow rate is considered.

Fig 2 shows the pressure profile for the fractional space diffusivity model at various values of the order, $\alpha$ at time $t = 12$ h. There is a decreasing response in the bottom hole pressure, $p_{wf}$ to the lowering effect of $\alpha$. This observation is characteristic of the proposed model for all simulations in time as $\alpha \to 0$. This is the sum effect of the fractional divergence, the built-in scale effect and the resistance.

Fig 3 shows the comparative example of the proposed model (FRDM) at $\alpha = 1$ and the classical diffusivity model (RDM) at time $t = 12, 24, 36$ and $48$ h. The results indicate as shown theoretical that there is a perfect correlation between the two models.

The decreasing variation of $\alpha$ registering a lowering response of $p_{wf}$ is analogous to that of a positive skin in petroleum reservoirs due to artificial drilling occurring few feet from the wellbore as shown in Fig 4.

A further consideration of the deviations of the proposed model is presented in Fig 5 with relatively strong and sharp response near the wellbore compared to the weak response further away from the wellbore. Table 2 shows a summary of statistics for the bottom hole response to $\alpha$.

The results indicate that although pressure drops are experienced in all cases where $\alpha < 1$, these pressure drops are registered more materially as $\alpha \to 0$. This phenomenon is absent in classical models analogous to the proposed model when $\alpha = 1$ in which case the model exhibits uniformity. For increasing levels of heterogeneity, the reservoir experiences more variability thus resulting in the pressure drop experienced near the wellbore.

## Conclusions

The proposed fractional space radial-cylindrical diffusivity model can be used to capture scale-dependent effects of real reservoirs which is important in applications. In addition, the proposed model exhibits fractional divergence, accounts for resistance to flow and can be used to account for the level of heterogeneity associated with each unique reservoir. The results are similar to the skin factor due to well damage. Heterogeneity has been attributed to the inhibition of fluids (Sharp et al. 2003) and thus results in lower borehole pressures than results from the classical model. The proposed model simulates the classical model and also can be applied in reservoirs with levels of non-uniformity which experiences resistance to fluid flow.

**Table 1** Relevant data

| Domain dimension | 2500 ft × 1° × 1 ft |
|---|---|
| Rock properties | $\phi = 20\%$ |
| | $k = 60$ md |
| Fluid properties | $B = 1.25$ bb/STB |
| | $\mu = 1.5$ cp |
| | $c_T = 1.2 \times 10^{-6}$ |
| Flow rate (constant) | 400 STB/day |

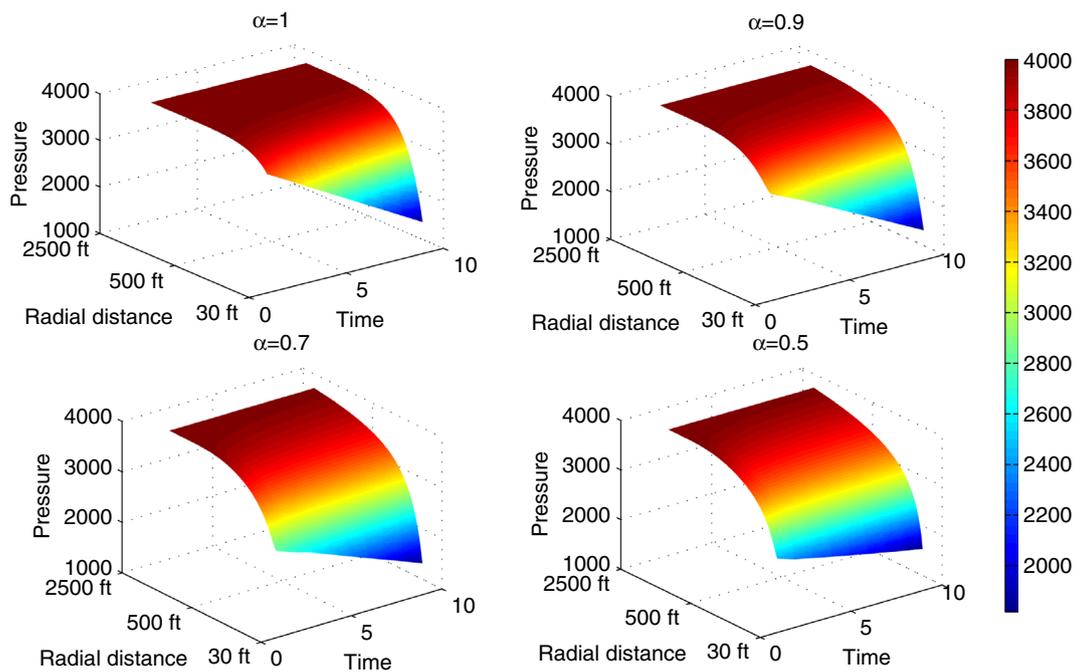

**Fig. 2** Effect of the variation of the level of heterogeneity

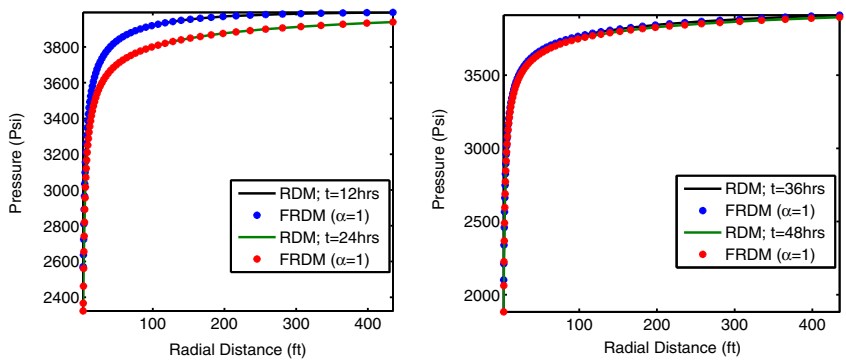

**Fig. 3** Comparison of the classical model with the proposed model

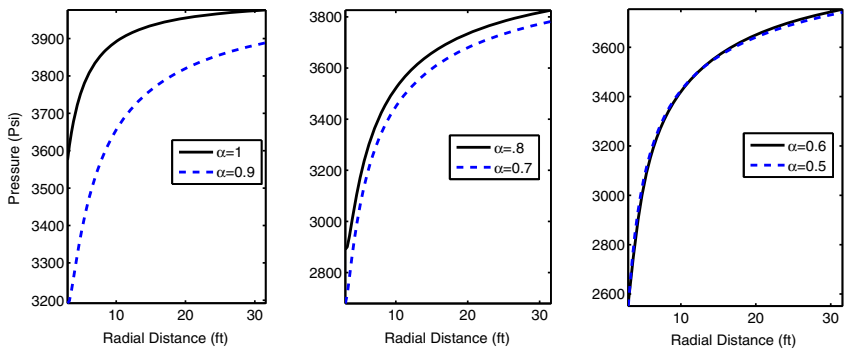

**Fig. 4** Fractional radial diffusivity model shows decreased $p_{wf}$ for variation in $\alpha$ at 12 h

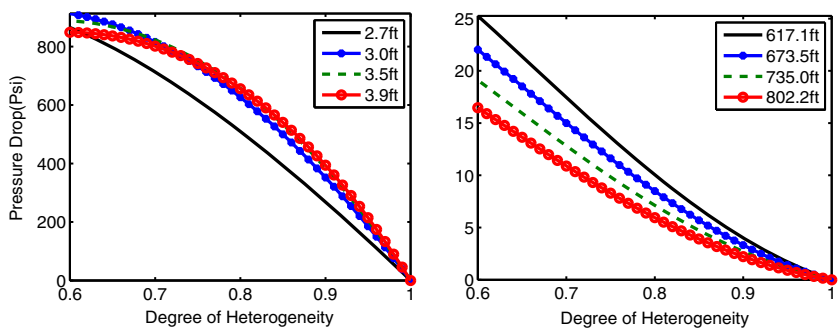

**Fig. 5** Fractional radial diffusivity model shows decreased $p_{wf}$ for variation in α at 12 h

**Table 2** Dependence of the level of heterogeneity, α and pressure drop

| Distance from wellbore | (ft) | Gradient | Intercept | Correlation coefficient |
|---|---|---|---|---|
| Well vicinity | 2.7 | $-2.1933 \times 10^{03}$ | $2.238 \times 10^{03}$ | $-0.9958$ |
|  | 3.0 | $-2.3004 \times 10^{03}$ | $2.4088 \times 10^{03}$ | $-0.9817$ |
|  | 3.2 | $-2.2062 \times 10^{03}$ | $2.3491 \times 10^{03}$ | $-0.9695$ |
|  | 3.5 | $-2.0785 \times 10^{03}$ | $2.3394 \times 10^{03}$ | $0.9601$ |
|  | 3.9 | $-1.9538 \times 10^{03}$ | $2.1232 \times 10^{03}$ | $0.9531$ |
| Further away | 617.1 | $-65.2254$ | $63.1375$ | $-0.9944$ |
|  | 673.5 | $-56.8070$ | $54.8273$ | $-0.9930$ |
|  | 735.0 | $-49.1707$ | $47.3166$ | $-0.9915$ |
|  | 802.2 | $-42.2969$ | $40.5880$ | $-0.9898$ |
|  | 875.5 | $-36.1584$ | $34.6046$ | $-0.9880$ |

**Acknowledgements** We would like to acknowledge the support received from the National Institute for Mathematical Sciences, Ghana, for this study.

**Compliance with ethical standards**

**Conflict of interest** The authors declare that they have no competing interests.